\documentclass[%
 reprint,
amsmath,amssymb,
aps,
prx,
]{revtex4-2}

\usepackage{graphicx}
\usepackage{dcolumn}
\usepackage{bm}
\usepackage{physics}

\usepackage{hyperref}
\usepackage[mathlines]{lineno}

\begin{document}
\renewcommand{\vb}{V\textsubscript{bias}~}
\preprint{APS/123-QED}

\title{STM observation of the hinge-states of bismuth nanocrystals}
\author{Tianzhen Zhang}
\affiliation{LPEM, ESPCI Paris, PSL Research University, CNRS, Sorbonne Universit\'e, 75005 Paris, France}

\author{Valeria Sheina}
\affiliation{Universités Paris-Saclay, CNRS, Centre de Nanosciences et de Nanotechnologies, 91120, Palaiseau, France}

\author{Sergio Vlaic}
\affiliation{LPEM, ESPCI Paris, PSL Research University, CNRS, Sorbonne Universit\'e, 75005 Paris, France}

\author{St\'ephane Pons}
\affiliation{LPEM, ESPCI Paris, PSL Research University, CNRS, Sorbonne Universit\'e, 75005 Paris, France}

\author{Dimitri Roditchev}
\affiliation{LPEM, ESPCI Paris, PSL Research University, CNRS, Sorbonne Universit\'e, 75005 Paris, France}

\author{Christophe David}
\affiliation{Universités Paris-Saclay, CNRS, Centre de Nanosciences et de Nanotechnologies, 91120, Palaiseau, France}

\author{Guillemin Rodary}
\affiliation{Universités Paris-Saclay, CNRS, Centre de Nanosciences et de Nanotechnologies, 91120, Palaiseau, France}

\author{Jean-Christophe Girard}
\affiliation{Universités Paris-Saclay, CNRS, Centre de Nanosciences et de Nanotechnologies, 91120, Palaiseau, France}

\author{Herv\'e Aubin}
\affiliation{Universités Paris-Saclay, CNRS, Centre de Nanosciences et de Nanotechnologies, 91120, Palaiseau, France}
\email{Herve.Aubin@universite-paris-saclay.fr} 


\date{\today}

\begin{abstract}
The recent application of topological quantum chemistry to rhombohedral bismuth established the non-trivial band structure of this material. This is a 2$^{nd}$order topological insulator characterized by the presence of topology-imposed hinge-states. The spatial distribution of hinge-states and the possible presence of additional symmetry-protected surface-states is expected to depend on the crystal shape and symmetries. To explore this issue, we have grown bismuth nanocrystals in the tens of nanometers on the $(110)$ surface of InAs. By scanning tunneling spectroscopy, we mapped the local density of states on all facets and identified the presence of the hinge-states at the intersection of all facets. Our study confirm the classification of bulk bismuth as a 2$^{nd}$order topological insulator. We propose that the ubiquitous presence of the hinge-states result from their tunnel-coupling across the nanometer-sized facets.
\end{abstract}

\maketitle


\section{Introduction}

Solids with filled electronic bands and no interactions may seem quite mundane, yet, band structures can have interesting non-trivial topological properties, which can be inferred from their symmetries.

To analyse the symmetry properties of an atom located on a lattice site, it is usual to employ the representation theory \cite{Tinkham_undated-wb} of point-groups to obtain the decomposition of the atomic orbitals as a direct sum of irreducible representations of the atomic site point-group symmetry. 

By analogy, to analyse the symmetry properties of the Bloch states arising from these orbitals, band representations theory, a representation theory of space groups \cite{Zak1982-ef,Cano2018-rg}, can be employed to obtain the decomposition of the band structure as a direct sum of irreducible band representations.

While band representation theory was barely used for decades, it have recently attracted intense interest with the demonstration that one band that cannot be decomposed into any linear combination of physical elementary band representation must have non-trivial topological properties. This gave birth to \emph{topological quantum chemistry} \cite{Bradlyn2017-ki} and enabled the identification and calculation of symmetry indicators for the 230 space groups \cite{Po2017-nn,Song2018-xi,Khalaf2018-re}, which are related to topological indexes, and facilitated large-scale search across databases to identify topological materials~\cite{Wieder2021-pm}. This has greatly expanded the topological classification of  band  insulators  beyond  the Z$_2$ topological insulators \cite{Hasan2010-vn,Qi2011-og} and revealed the existence of numerous symmetry-indicated topological phases such as mirror-symmetry-protected topological crystalline insulators (TCIs) \cite{Teo2008-yk,Fu2011-lj}, rotational-symmetry–protected  TCIs \cite{Fang2019-ur} and higher-order topological insulators (HOTIs) \cite{Song2017-jj,Schindler2018-ro,Khalaf2018-dp,Schindler2018-sl,Wang2019-jx}.
 
Although the lowest-symmetry HOTI models exhibit gapped two-dimensional (2D) surfaces-states and gapless one-dimensional (1D) hinge-states \cite{Wang2019-jx,Po2017-nn,Song2018-xi,Khalaf2018-re}, typical solid-state HOTIs have additional crystal symmetries that can protect gapless surface-states; thus, a HOTI can coexist with a TCI. Depending on the sample termination, the same crystal  may exhibit either 2D surface-states or gapped facets separated by topological hinge-states. 


The application of topological quantum chemistry to rhombohedral bismuth (Bi), with space-group  R$\bar{3}$m (\#166) and point-group D\textsubscript{3d} has shown that its valence and conduction bands result from a split elementary band representation\cite{Schindler2018-sl}, which implies non-trivial topological properties. These topological properties result from a double band inversion
\cite{Hsieh2008-gk,Teo2008-yk} at the same time-reversal-invariant momenta (TRIM) points. The topological phase can be described as two superposed copies of a TCI protected by time reversal T, threefold rotation C$_{3[111]}$ about the $[111]$ direction and inversion I symmetries \cite{Schindler2018-sl}, characterized by the Z$_4 = 2$ topological index \cite{Schindler2018-ro,Wang2019-jx,Po2017-nn,Song2018-xi,Khalaf2018-re}. The fourfold Dirac surface-states of this topological phase are unstable\cite{Wang2019-jx,Wieder2018-sz} and it has been shown that a crystal with an inversion-symmetric shape, such as a rod of hexagonal section oriented along the $[111]$ direction, the mass of the Dirac states is of opposite sign on surfaces with inversion - related  Miller indices, resulting in helical hinge modes encircling the crystal \cite{Schindler2018-sl}.
\begin{figure*}[ht]
	\includegraphics[width=0.95\linewidth]{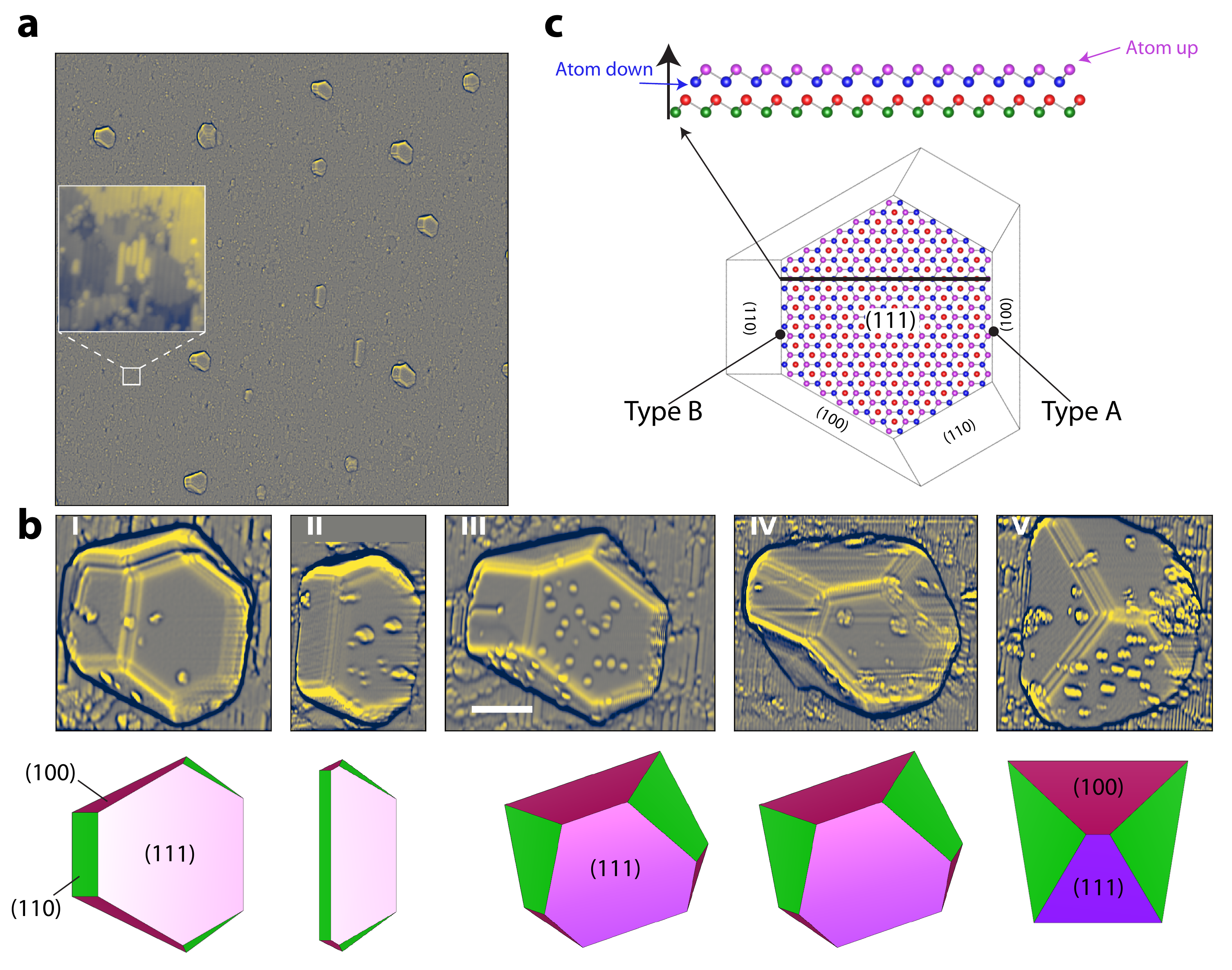}
	\centering
	\caption{\label{fig:topo} (a) 800 nm $\times$ 800 nm Laplacian $\Delta_{xy}z(x,y)$ transform of a topographic STM image (I\textsubscript{set} = 30 pA, \vb = 1.5 V). The inset is a zoom on a zone 30 nm $\times$ 30 nm.  (b) Laplacian $\Delta_{xy}z(x,y)$ transforms of selected topographic images, labelled from I to V. The scale bar on the third image is 10 nm long. Below each image, a model of the shape is provided where are indicated the Miller indices of the facets. (c) Top : Atomic model of two atomic bilayers of Bi. Bottom : Atomic model of a NC. The type B edge, last atom of the first bilayer pointing down, is located at the hinge between the $(111)$ and $(110)$ facets. The type A edge, last atom of the first bilayer pointing up, is located at the hinge between the $(111)$ and $(100)$ facets.}
	\centering
\end{figure*}

Experimentally, on hexagonal cavities in the $(111)$ surface of \emph{bulk} Bi \cite{Drozdov2014-qi}, scanning tunneling microscopy (STM) and spectroscopy (STS) identified edge-states on every two edges of the cavity. Because two-dimensional bismuth was predicted theoretically to be a two-dimensional quantum spin Hall insulator (QSHI) with Kane-Mele index Z$_2 = 1$ \cite{Murakami2006-cb,Wada2011-jq}, the edge-states were identified with the topological states of a QSHI altered by the coupling to the bulk\cite{Drozdov2014-qi}. While later STM works indeed confirmed the presence of topological edge-states on thin bismuth films\cite{Peng2018-fl,Yang2017-tf} and bismuthene\cite{Reis2017-kl,Stuhler2022-yd,Salehitaleghani2022-qz}, those observed on \emph{bulk} Bi cavities were later reinterpreted as the signature of the hinge-states of a HOTI\cite{Schindler2018-sl}. Interestingly, the edge-states in bismuthene and hinge-states in Bi are related as it was shown that the stacking of multiple 2D QSHI leads either to a weak 3D topological insulator, as for example Pt\textsubscript{2}HgSe\textsubscript{30} \cite{Marrazzo2018-qz,Facio2019-ap,Marrazzo2020-zv,Cucchi2020-ev} or a HOTI, as Bi\cite{Schindler2018-sl}, WTe\textsubscript{2}\cite{Wang2019-jx} and Bi\textsubscript{4}Br\textsubscript{4} \cite{Noguchi2021-ez}. Finally, the presence of the hinge-states was also inferred from SQUID-like supercurrent oscillations observed with Josephson circuits fabricated on Bi\cite{Schindler2018-sl,Bernard2021-bs} and WTe\textsubscript{2} \cite{Choi2020-cs,Kononov2020-lq,Huang2020-cc}.

In addition to the HOTI phase protected by the C$_{3[111]}$ symmetry, it has been shown\cite{Hsu2019-yf} that Bi is also host of a first-order TCI state protected by a twofold rotational symmetry C$_{2[1\bar{1}0]}$ about the $[1\bar{1}0]$ direction, resulting in a pair of topological Dirac surface-states on its $(1\bar{1}0)$ surface, coexsisting with the hinge-states of the HOTI phase. On $(110)$ oriented Bi films, recent STM/STS measurements\cite{Aggarwal2021-wl} have identified  hinge-states that could be described by a HOTI protected by time-reversal symmetry and bulk two-fold rotation C$_{2[1\bar{1}0]}$\cite{Hsu2019-yf}. The surface-states were not observed though, because of the broken translation symmetry on the $(1\bar{1}0)$ facets.

These recent theoretical and experimental works suggest that the distribution of surface- and hinge-states on a Bi crystal depends on its shape and transformation properties under C$_{3[111]}$, C$_{2[1\bar{1}0]}$ and inversion symmetries.

To explore the relationship between surface-states, hinge-states and the crystal shape, we have grown Bi nanocrystals (NCs) on the $(110)$ surface of InAs and characterized its electronic properties by STM/STS. Thanks to the small size of the NCs and using very sharp STM-tips, we have been able to perform the STM/STS characterization of multiple facets and hinges of the NCs. We clearly identified hinge-states, confirming qualitatively the classification of Bi as a HOTI. However, we found that the spatial distribution of the density of states due to hinge-states could not be reproduced easily by a simple model with hinge-states appearing at the intersection of facets with Dirac masses of opposite signs\cite{Schindler2018-sl,Hsu2019-yf}. Instead we found hinge-states at the intersection of all facets, provided they are sufficiently extended. We suggest that the ubiquitous presence of hinge-states is the consequence of their tunnel-coupling across the nanometer-sized facets.

\section{Nanocrystal growth}

The Bi NCs have been grown on the $(110)$ surface of n-doped InAs, which substrate has been cleaved in ultra-high vacuum at base pressure P$~\approx~10^{-10}$ mbar. The $(110)$-surface of InAs substrates has also been employed previously for the growth of superconducting Pb NCs \cite{Vlaic2017-dh,Zhang2018-pi} and Bi nanolines\cite{Schmidt1996-dl,Betti1999-vj}. By thermal evaporation, we deposited a nominal quantity of three monolayers of Bi at a temperature of 500 K. This is the optimum temperature for the growth of nanolines\cite{Betti1999-vj}, where we checked their formation by low energy electron diffraction (LEED). A topographic STM image is shown in Fig.~\ref{fig:topo}a, where Bi NCs are visible as well as nanolines in the zoom insert. To accentuate the facets and edges, a Laplacian filter $\Delta_{xy}z(x,y)$ is applied to all topographic images. Fig.~\ref{fig:topo}b show topographic images zoomed on five selected Bi nanocrysals, labelled I to V. The lateral size of NCs is between 10 nm and 40 nm, the height is between 5 nm and 15 nm. To obtain such topographic images free from artifacts resulting from the shape of the tip, a tip-shaping procedure has been employed where the tip is approached into contact with the InAs surface, then, a large current of 1 $\mu A$ is injected and the tip is slowly retracted. This tip-elongation procedure is repeated until no effects of the shape of the tip is visible on the topography of the NCs. Using the crystallographic software VESTA$^@$ and employing the crystal structure of Bi, we model the shape of all NCs and show the results below the corresponding NCs in Fig.~\ref{fig:topo}b. From these Wulf constructions, the crystallographic planes $(111)$, $(110)$ and $(100)$ can be clearly identified for each NC. 

On the hexagonal cavity studied on the $(111)$ surface of bulk Bi \cite{Drozdov2014-qi}, two type of edges were identified with edge A (B) corresponding to the edge with last Bi atom pointing up (down). We use the same notation. As shown by the atomic model in Fig.~\ref{fig:topo}c, the location of edges A and B can be identified on NCs, which was not possible on the cavity. Edge A (B) is located at the intersection, i.e. hinge, between the $(111)$ and the $(100)$ ( $(110)$ ) facets. 

\section{Scanning Tunneling Spectroscopy}

\begin{figure*}
	\includegraphics[width=0.9\linewidth]{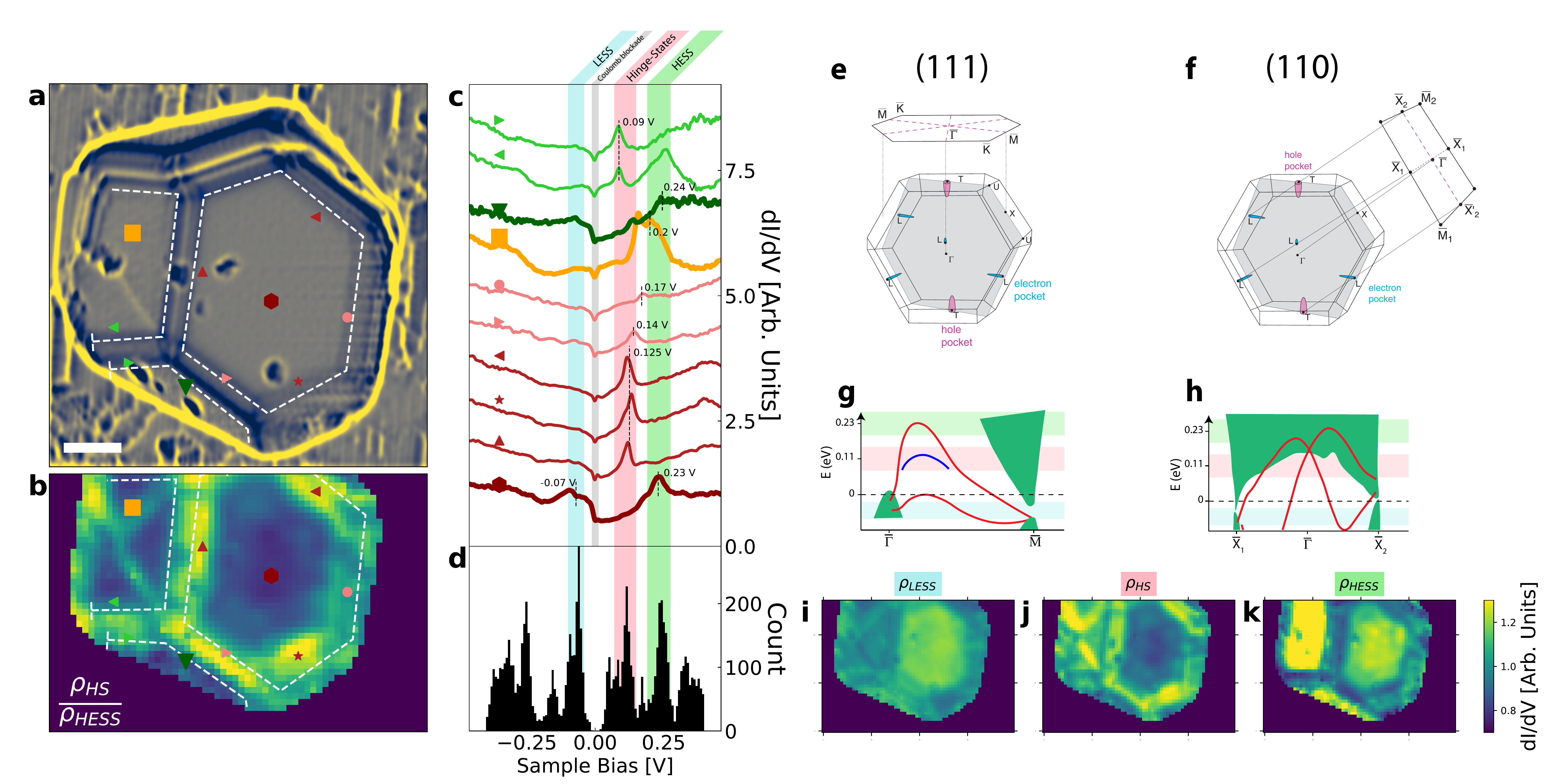}
	\centering
	\caption{STS data on NC I. (a) Laplacian $\Delta_{xy}z(x,y)$ transform of a topographic STM image (I\textsubscript{set} = 30 pA, \vb = 1.5 V). The scale bar is 5 nm long. (b) Conductance map $\frac{\rho_{HS}}{\rho_{HESS}}$ obtained by taking the ratio of the DC integrated on energy range of hinge-states, $\rho_{HS}$, with the DC integrated on energy range of HESS, $\rho_{HESS}$. The scale is the same as panel (a). (c) DC for selected points indicated as colored symbols on panels ab. The vertical colored rectangles indicate the energy ranges of interest corresponding, respectively, to LESS, Coulomb gap, hinge-states and HESS. The vertical dashed lines indicates conductance peaks of interest. See text. (d) Histogram of the energy values of conductance peaks identified in all DC curves measured on the NC. (e) Bulk Brillouin zone and pseudo-Brillouin zone of the $(111)$ facet. (f) Bulk Brillouin zone and pseudo-Brillouin zone of the $(110)$ facet. (g) Extracted from Ref. \cite{Hofmann2006-al}, the relations dispersions of projected bulk states (green areas),  surface-states (red lines) and hinge-states (blue line) on the $(111)$ facet. (h) Extracted from Ref. \cite{Hofmann2006-al}, the relations dispersions of projected bulk states (green areas) and surface-states (red lines) on the $(110)$ facet. (i), (j) and (k) Conductance maps $\rho_{LESS}$, $\rho_{HS}$ and $\rho_{HESS}$ obtained by integrating the DCs on energy range of LESS, hinge-states and HESS, respectively.}
	\centering
    \label{fig:nanoI}
\end{figure*}

\begin{figure*}
	\includegraphics[width=0.9\linewidth]{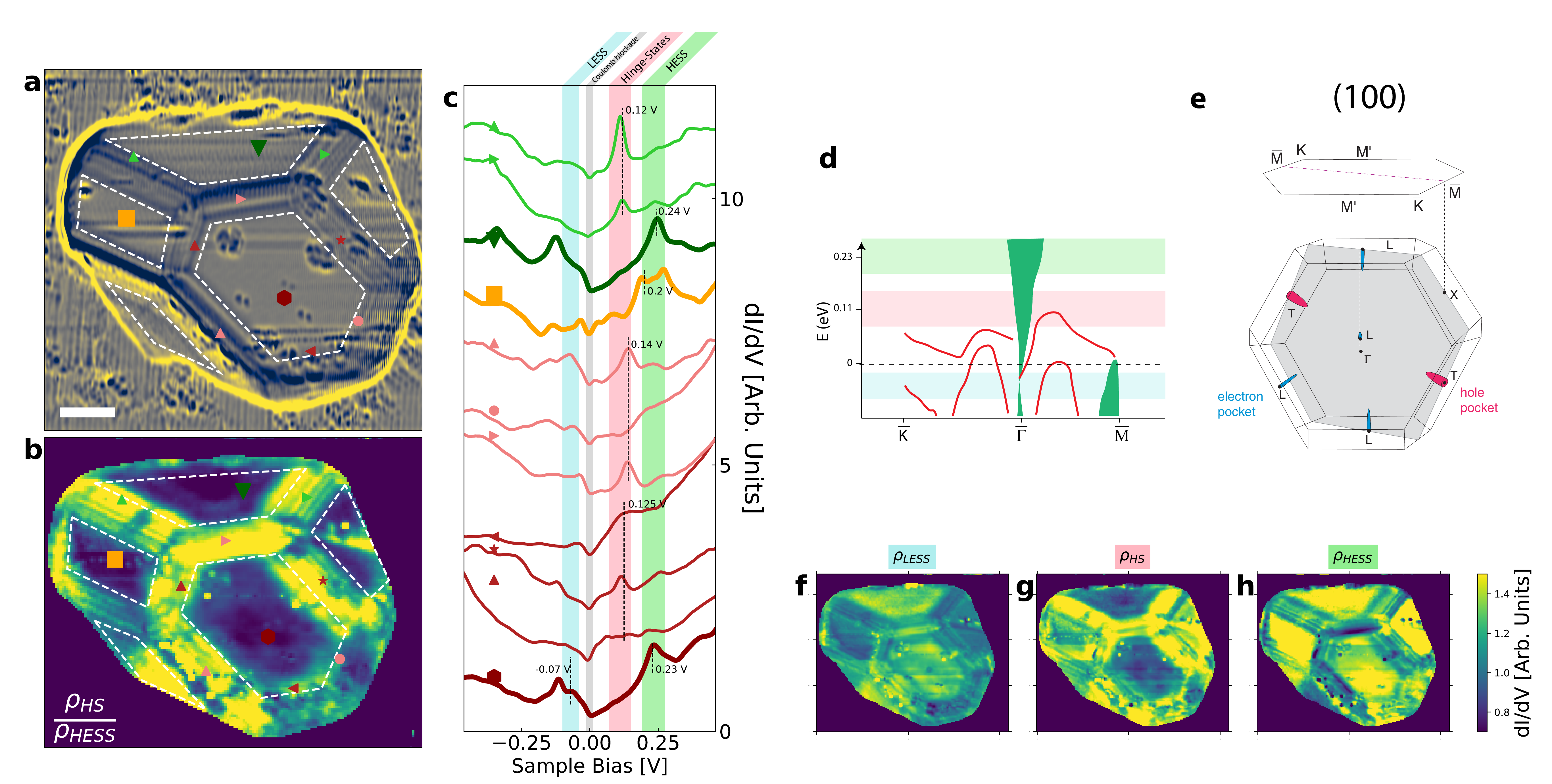}
	\centering
	\caption{STS data on NC IV. (a) Laplacian $\Delta_{xy}z(x,y)$ transform of a topographic STM image (I\textsubscript{set} = 30 pA, \vb = 1.5 V). The scale bar is 5 nm long. (b) Conductance map $\frac{\rho_{HS}}{\rho_{HESS}}$ obtained by taking the ratio of the DC integrated on energy range of hinge-states, $\rho_{HS}$, with the DC integrated on energy range of HESS, $\rho_{HESS}$. The scale is the same as panel (a). (c) DC for selected points indicated as colored symbols on panels ab. The vertical colored rectangles indicate the energy ranges of interest corresponding, respectively, to LESS, Coulomb gap, hinge-states and HESS. The vertical dashed lines indicates conductance peaks of interest. See text. (d) Bulk Brillouin zone and pseudo-Brillouin zone of the $(100)$ facet. (e) Extracted from Ref. \cite{Hofmann2006-al}, the relations dispersions of projected bulk states (green areas),  surface-states (red lines) and hinge-states (blue line) on the $(100)$ facet. (f), (g) and (h) Conductance maps $\rho_{LESS}$, $\rho_{HS}$ and $\rho_{HESS}$ obtained by integrating the DCs on energy range of LESS, hinge-states and HESS, respectively.}
	\centering
	\label{fig:nanoIV}
\end{figure*}

\begin{figure*}
	\includegraphics[width=0.95\linewidth]{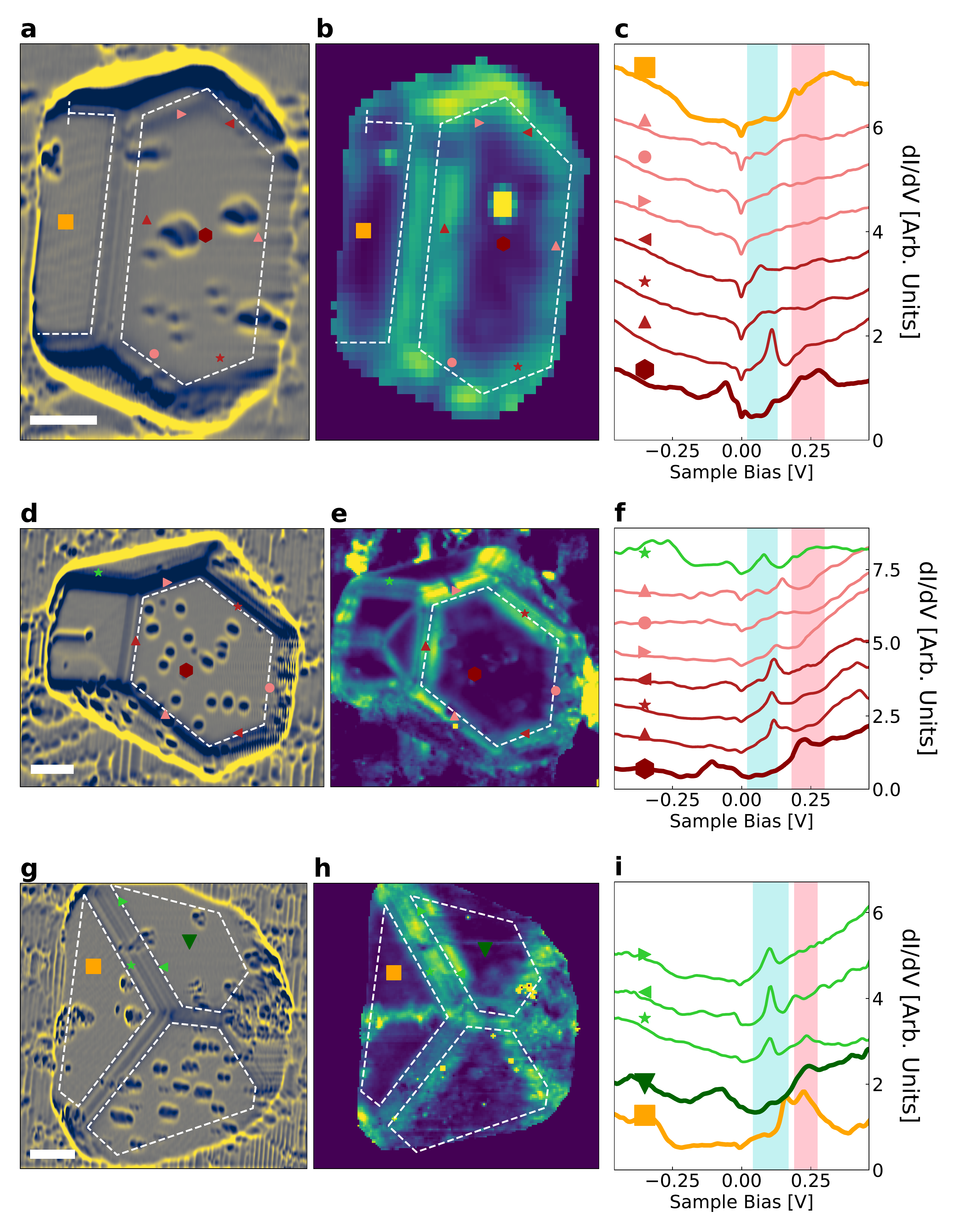}
	\centering
	\caption{ (abc) STM/STS data for NC II. (a) Laplacian $\Delta_{xy}z(x,y)$ transform of a topographic STM image (I\textsubscript{set} = 30 pA, \vb = 1.5 V). The scale bar is 5 nm long. (b) Conductance map $\frac{\rho_{HS}}{\rho_{HESS}}$ obtained by taking the ratio of the DC integrated on energy range of hinge-states, $\rho_{HS}$, with the DC integrated on energy range of HESS, $\rho_{HESS}$. The scale is the same as panel (a). (c) DC for selected points indicated as colored symbols on panels ab. The vertical colored rectangles indicate the energy ranges of interest corresponding, respectively, to the hinge-states and HESS. (def) STM/STS data for NC III. (ghi) STM/STS data for NC V.}
	\centering
	\label{fig:nanosupp}
\end{figure*}

On these 3D NCs, it is possible to perform STS on multiple facets. We will now detail the results obtained on two selected NCs. The results for NC I, which has large $(111)$ and $(110)$ facets, are shown in Fig.~\ref{fig:nanoI}. The results for NC IV,  which has large $(111)$ and $(100)$ facets, are shown in Fig.~\ref{fig:nanoIV}. Additional STS results for the NCs II, III and V are shown in Fig.~\ref{fig:nanosupp}. 

For all NCs, a small Coulomb gap about 25 meV is visible at zero bias, as highlighted in Fig.~\ref{fig:nanoI}c and Fig.~\ref{fig:nanoIV}c. As described in the previous work on Pb NCs\cite{Vlaic2017-dh,Zhang2018-pi}, when a NC is grown atop a 2D electron gas of Fermi wavelength larger than the NC lateral sizes, a quantum constriction effect occurs, similar to what observed in quantum-point contacts, reducing the transmission to values smaller than one quantum of conductance. In this situation, the NC is only weakly coupled to the substrate and so Coulomb blockade occurs. In our system, the Bi nanowires and NCs have expected Fermi wavelengths in the 10 - 30 nm range \cite{Hofmann2006-al}, comparable to NCs lateral sizes, explaining the observation of Coulomb blockade and implying that the Bi NCs are only weakly coupled to the substrate.

Fig.~\ref{fig:nanoI}c and Fig.~\ref{fig:nanoIV}c show the differential conductance (DC) $\dv{I}{V}{(V)}$ spectra measured on NC I and NC IV, respectively, at positions indicated by colors symbols on corresponding panels ab. 

In the middle of the $(111)$ facets (red hexagon symbols), the DCs present conductance peaks at \vb = 0.23 V. This is a consequence of the saddle point in the higher-energy relation dispersion of $(111)$ surface-states (HESS)\cite{Hofmann2006-al,Koroteev2004-hb,Koroteev2008-hw}, shown in  Fig.~\ref{fig:nanoI}g, along the $\mathbf{\bar{\Gamma}} - \mathbf{\bar{M}}$ direction of the Brillouin zone shown in Fig.~\ref{fig:nanoI}e. The corresponding energy range is indicated as green zones in Fig.~\ref{fig:nanoI}cdgh and Fig.~\ref{fig:nanoIV}cd. The DCs also show a sharp increase near \vb = -0.01 V, which is a consequence of the saddle point in the lower-energy relation dispersion of $(111)$ surface-states, followed by a maxima about \vb = -0.07 V, identified as low-energy surface-states (LESS) where the corresponding energy range is indicated as blue zones in Fig.~\ref{fig:nanoI}cdgh and Fig.~\ref{fig:nanoIV}cd. 

On the edges of the $(111)$ facets, the conductance peak at \vb = 0.23 V due to HESS is missing, instead, peaks in the DCs are observed at lower energy. For NC I, conductance peaks are observed at \vb = 0.125 V on all type-B edges (red symbols). On the two type-A edges measured (due to STM drift, the third one has not been measured), one large conductance peak is visible at \vb = 0.14 V on the edge (pink right triangle)  where the $(100)$ facet is well defined  but only a very small conductance peak (\vb = 0.17 V) is visible on the second one (pink disk); we suggest this is due to the absence of the $(100)$ facet for this edge where the $(111)$ facet is reaching the InAs surface. For NC IV, on the type-B edges, the conductance peak at \vb = 0.125 V is well defined only on the longest edge (red star symbol) but not on the two other short edges (red left- and up-triangles). On the type-A edges, the conductance peak at \vb = 0.14 V is visible on two edges (pink right- and up-triangles), where the $(100)$ facet is well defined, but not on the third one (pink disk symbol) where the $(100)$ facet is absent and the $(111)$ facet is reaching the InAs surface. These conductance peaks on both type of edges are identified as the one-dimensional van Hove singularities (vHS) of hinge-states. Their energy range is indicated as pink zones in Fig.~\ref{fig:nanoI}cdgh and Fig.~\ref{fig:nanoIV}cd. Thus, in contrast to the observation on cavities in the $(111)$ surface of bulk Bi crystal\cite{Drozdov2014-qi}, we find that the hinge-states vHs conductance peaks are visible on both type of edges. As is also seen on the other NCs, Fig.~\ref{fig:nanosupp}, the conductance peaks are visible only when the two facets making the hinge are extended and well defined.

In the middle of the $(110)$ facet (orange square symbols), the DCs feature broad maxima at \vb = 0.18 V, consistent with the observation of Ref.~\cite{Aggarwal2021-wl}. This maxima in the density of states is due to $(110)$ surface-states\cite{Hofmann2006-al,Takayama2014-to,Takayama2015-za}, whose dispersion relation is shown Fig.~\ref{fig:nanoI}h along the $\mathbf{\bar{X}_1} - \mathbf{\bar{\Gamma}} - \mathbf{\bar{X}_2}$ direction of the Brillouin zone shown in Fig.~\ref{fig:nanoI}f. 

In the middle of the $(100)$ facets (green down-triangle symbols), mostly visible in Fig.~\ref{fig:nanoIV}c of NC IV, the DCs feature broad maxima at \vb = 0.24 V, which is likely due to $(111)$ surface-states\cite{Hofmann2006-al}. 

Finally, on the hinges between the $(110)$ and $(100)$ facets (light green symbols), the DCs feature conductance peaks at \vb $\approx$ 0.1 V that indicates the presence of hinge-states. This is remarkable as they do not involve the $(111)$ facet, which confirms the classification of Bi as a HOTI. 

Two comments are in order. 

First, we note that the energy of the vHs slightly changes from one hinge to the other; in particular, they have energies $\approx$ 0.125 eV for type-B edges and $\approx$ 0.14 eV for type-A edges. This changes of energy could be explained by the tunnel-coupling between hinge-states.  
 
Second, on all edges, the vHs conductance peaks appear concomitantly with the disappearance of the conductance peaks due to surface-states on all three facets. Such an anti-correlation between the HESS conductance peak on the $(111)$ surface and the hinge-states vHs conductance peaks was also observed previously\cite{Drozdov2014-qi}.

To see this more clearly, we acquired full conductance maps, typically on a spatial grid of 64 pixels $\times$ 64 pixels and 512 voltage points.  Fig.~\ref{fig:nanoI}d shows an histogram of the energies of the conductance peaks identified in the 4096 DC spectra measured on the NC I. This histogram shows that the LESS, HESS and hinge-states are at the origin of all conductance peaks observed in the energy range [-0.25 eV, 0.25 eV]. There is no other major spectral feature of interest in this energy range on the whole NC.

To obtain maps of the local density of states, we remove the effects of changing tip-height and tunnel barrier energy by normalizing the DC spectra assuming that the total density of states is conserved on the energy range [-0.5 V, 0.5 V] measured. That is, the DC $\dv{I}{V}{(V)}$ spectra are normalized by their integrated values $\int_{-0.5}^{0.5}{\dv{I}{V}{}dV}$. Then, the normalized spectra are integrated on the three energy ranges corresponding to HESS [0.19 V, 0.27 V], LESS [-0.1 V, -0.04 V] and hinge-states [0.07 V, 0.15 V], to give $\rho_{HESS}$, $\rho_{LESS}$ and $\rho_{HS}$, respectively, shown Fig.~\ref{fig:nanoI}ijk and Fig.~\ref{fig:nanoIV}fgh for NC I and NC IV, respectively.

At the energy of HESS, Fig.~\ref{fig:nanoI}k and Fig.~\ref{fig:nanoIV}h, a large density of states is visible on all facets except on the edges. In contrast, at the energy of hinge-states vHs, Fig.~\ref{fig:nanoI}j and  Fig.~\ref{fig:nanoIV}g, a large density of states is observed on almost all hinges, between the $(111)$ facet and $(110)$, $(100)$ facets as well as between the $(110)$ facets and the $(100)$ facets. Note that a large density of hinge-states is also observed in the middle of the largest $(110)$ facet of NC I, seemingly as a consequence of the defect observed in the topography, Fig.~\ref{fig:nanoI}a. Note also that the density of states due to LESS is mostly visible on the $(111)$ and $(100)$ facets and not on the $(110)$ facets. Looking at the conductance maps Fig.~\ref{fig:nanoI}jk and Fig.~\ref{fig:nanoIV}gh for both NC I and NC IV, respectively, the anti-correlation between the density of surface-states, $\rho_{HESS}$, and the density of hinge-states in the vHs, $\rho_{HS}$, is obvious. This suggests that a map of hinge-states can be obtained by calculating the ratio $\rho_{HS}/\rho_{HESS}$, which is shown Fig.~\ref{fig:nanoI}b and Fig.~\ref{fig:nanoIV}b for both NC I and NC IV, respectively. In addition, Fig.~\ref{fig:nanosupp} show the topography, ratio $\rho_{HS}/\rho_{HESS}$ and DC spectra for the three additional NCs : II, III and V. 

For all NCs, the map $\rho_{HS}/\rho_{HESS}$ show that the topological states are clearly visible only at the hinges where two distinct Bi facets are well defined. See how the conductance peaks due to hinge-states disappear where the $(111)$ facet reaches the InAs surface. This suggests that the hinge-states indeed arise from the interaction between surface-states as expected theoretically. More precisely, the theoretical model proposed in Ref.~\cite{Schindler2018-sl} implies the existence of two sets of surface-states resulting from two copies of the 3D Z$_2$ topological insulator. The off-diagonal coupling between the two sets opens a gap for the surface-states, leaving only in-gap hinge-states at the edges of the facets. Thus, our STM/STS study of NCs confirm the existence of hinge-states and so the classification of Bi as a HOTI. 

\begin{figure*}
	\includegraphics[width=0.95\linewidth]{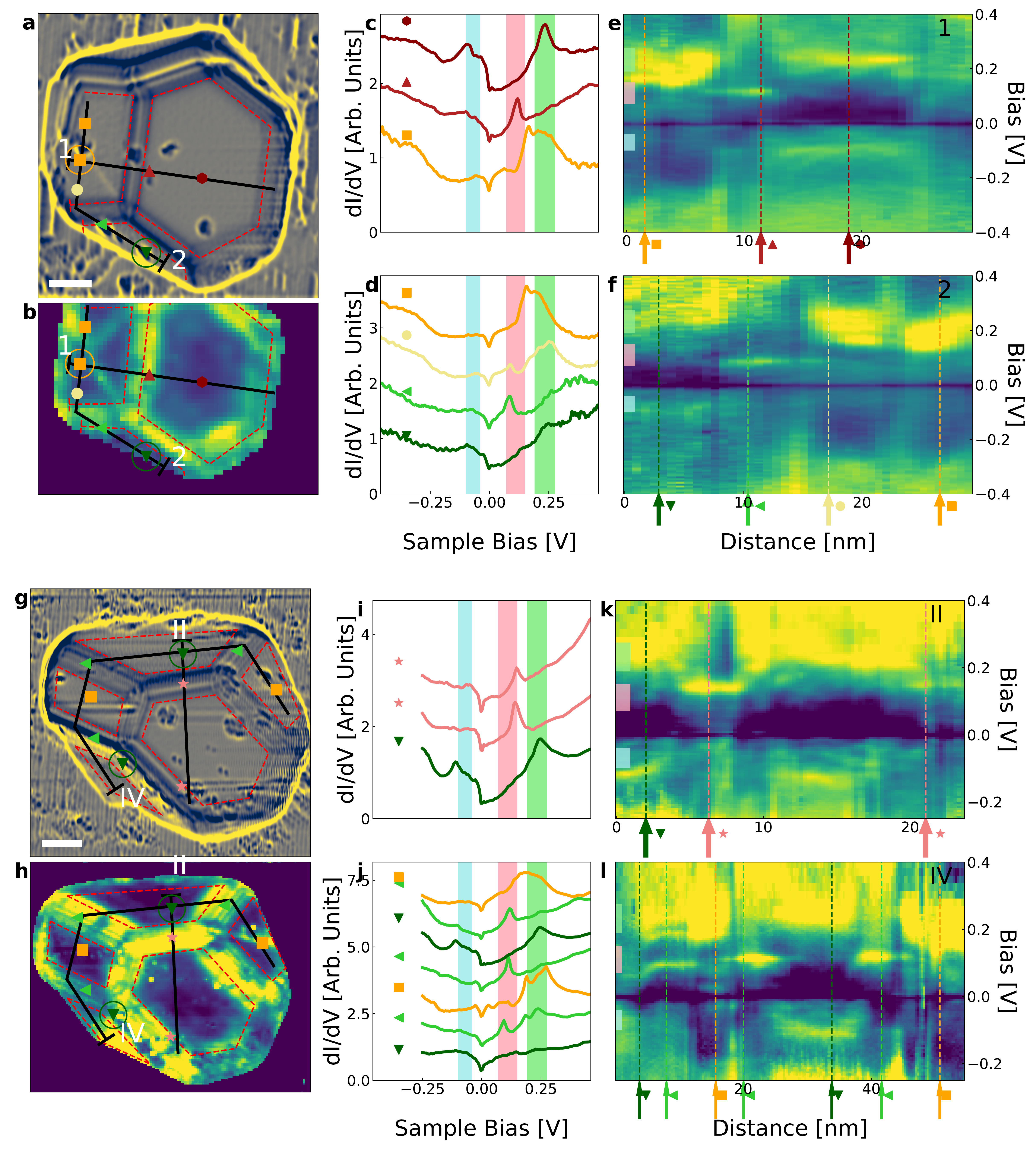}
	\centering
		\caption{STM/STS data for NC I. (a) Laplacian $\Delta_{xy}z(x,y)$ transform of a topographic STM image (I\textsubscript{set} = 30 pA, \vb = 1.5 V). The scale bar is 5 nm long. (b) Conductance map $\frac{\rho_{HS}}{\rho_{HESS}}$ obtained by taking the ratio of the DC integrated on energy range of hinge-states, $\rho_{HS}$, with the DC integrated on energy range of HESS, $\rho_{HESS}$. The scale is the same as panel (a). (cd) DC for selected points indicated as colored symbols on panels ab and indicated as vertical dashed lines on panels ef. (ef) Conductance maps as function of sample bias and distance along the profiles drawn as black lines on panels ab. (ghijkl) STM/STS data for NC IV.}
	\centering
	\label{fig:nanocut}
\end{figure*}

\section{Discussion}

Despite global consistency of the STS data with a HOTI model, there are two features that remain be understood theoretically. First, as mentioned above, the spectral weight associated with the hinge-states in the vHs appears concomitantly with the disappearance of the spectral weight associated with the HESS. To highlight this anti-correlation between HESS and hinge-states, Fig.~\ref{fig:nanocut}efkl show the DC maps as function of sample voltage and distance along profiles drawn on the topographic images Fig.~\ref{fig:nanocut}a(g) and ratio maps Fig.~\ref{fig:nanocut}b(h) for NC I (IV). These maps show that the hinge-states appear at the expense of the surface-states on all facets.
In the model of Ref.~\cite{Schindler2018-sl}, this is not expected, the gap should remain constant on the whole surface of the NC, including the hinges, and so should the spectral weight due to surface-states. Second, according to this model applied to a rod of hexagonal section\cite{Schindler2018-sl}, the hinge-states should not exist on all hinges but only on those separating facets with Dirac masses of opposite signs. Instead, we observe topological states on all hinges except where the facet reaches the InAs substrate. We suggest that this ubiquitous presence result from the coupling between hinge-states on opposite sides of the facets. Indeed, it is well established that the tunnel coupling between the edge-states at opposite edges (surfaces) of a 2D (3D) topological insulator leads to a gap opening in the relation dispersion of the edge-states Dirac spectrum \cite{Jung2021-dk,Stuhler2022-yd}. This is also true for the tunnel-coupling between edge-states of a second-order topological insulator as shown recently for the corner-states of a 2D second-order topological insulator \cite{Arai2021-og}. In this last theoretical work, it has been shown that for well \emph{localized} corner-states, the corresponding density of states is large on only some of the corners and not on others, however, when the wavefunctions of corner-states overlap, the corresponding density of states become equally distributed on all corners. In our small NCs, significant coupling between hinge-states is to be expected and could explain their presence on all hinges as well as the slightly different energy position of the vHs observed for different hinges. Furthermore, at hinges where several atomic steps exists, individual hinge-states separated in space by terraces should exist and for an even number of such states, they can hybridize and gap out. Note also that the observation of a vHS conductance peak does not imply that the hinge-state is topological. It is possible that only some of the hinge-states are gapless, while others are gapped, however, on the background of all the conducting states such a gap is hard to observe by STM.

\section{Conclusion}

To summarize, we have shown that Bi NCs could be grown on the $(110)$ InAs surface. Their size in the tens of nanometer make them suitable for STM/STS studies on all facets of the tri-dimensional NC. We identified the hinge-states vHs at all hinges between the three facets $(111)$, $(110)$ and $(100)$, however, on edges where the facet ends on the InAs substrate, no vHs is observed. These observations are consistent with the classification of Bi as a HOTI. The ubiquitous observation of the vHs on all hinges suggests that the hinge-states are tunnel-coupled because of the small size of the NCs.

\begin{acknowledgments}
We acknowledge financial support from ANR MECHASPIN Grant No. ANR-17-CE24-0024-02 and ANR FRONTAL Grant No. ANR-19-CE09-0017-02. We thank S. Guéron, H. Bouchiat, T. Neupert and M. Denner for discussions, reading of the manuscript and suggestions.
\end{acknowledgments}

\bibliography{biblio.bib}

\end{document}